\documentstyle[12pt]{article}
\begin{document}
\title{{\bf ATTACHING THEORIES OF CONSCIOUSNESS
TO BOHMIAN QUANTUM MECHANICS}
\thanks{Alberta-Thy-12-95, quant-ph/9507006, to be published in {\em Bohmian
Quantum Mechanics and Quantum Theory:  An Appraisal}, edited by J. T. Cushing,
A. Fine, and S. Goldstein (Kluwer, Dordrecht, 1996)}}
\author{
Don N. Page
\thanks{Internet address:
don@phys.ualberta.ca}
\\
CIAR Cosmology Program, Institute for Theoretical Physics\\
Department of Physics, University of Alberta\\
Edmonton, Alberta, Canada T6G 2J1
}
\date{(1995 June 30)}

\maketitle
\large
\begin{abstract}
\baselineskip 14 pt

The de Broglie-Bohm theory of quantum
mechanics (here simply called Bohmian Mechanics or BM)
[1-10]
is an augmentation of ``bare'' quantum mechanics (the bare theory being given
by an algebra of operators and a quantum state that sets the expectation values
of these operators) that includes a definite history or Bohmian trajectory.
This definite trajectory gives BM a somewhat more classical flavor than most
other forms of quantum mechanics (QM) (though the trajectory certainly has
highly nonlocal and other nonclassical aspects in its evolution), but to see
whether or not this makes a difference for observations by conscious beings,
one needs to attach theories of conscious perceptions to BM and other forms of
QM.  Here I shall propose various forms of theories of consciousness for BM,
which I shall call {\it Sensible Bohmian Mechanics} (SBM), and compare them
with a proposal I have made for a theory of consciousness attached to bare QM,
which I call {\it Sensible Quantum Mechanics} (SQM)
[11-15].
I find that only certain special forms of SBM would give essentially similar
predictions as SQM, though a wider class might be in practice indistinguishable
to any single observer.  I also remain sceptical that a viable complete form of
SBM will turn out to be as simple a description of the universe as a viable
complete form of SQM, but of course it is too early to know yet what the form
of the simplest complete theory of our universe is.
\\
\\
\end{abstract}
\normalsize
\baselineskip 14.7 pt

I should explain at the outset that my attempt to incorporate consciousness
within physics is nonstandard, since most physicists would probably consider
this attempt premature, or simply not part of physics.  I agree that it {\it
is} probably premature to try to give a complete detailed theory of
consciousness, but since we have developed mathematical frameworks like
mechanics for describing other aspects of the universe, even though we do not
yet know the correct complete detailed form of the mechanics describing our
universe, it surely would be helpful to try to develop a mathematical framework
for describing consciousness, even though we are a very long way from a correct
complete detailed theory for it.  The objection that consciousness is not part
of physics may be historically valid as a sociological analysis of what most
physicists actually study (for the reason, I would guess, that physics
concentrates on the simplest fundamental descriptions of aspects of our
universe, and so far consciousness does not seem to be simple enough to be
included in physics).  However, physics has historically continued to be
extended to describe a broader and broader range of aspects of our universe, so
that in some branches of physics there is even talk today of `theories of
everything.'  Thus it seems plausible to me that physics should eventually
attempt to describe consciousness itself.  Certainly talk of `theories of
everything' in physics seems rather hollow if physics is required to refuse to
consider consciousness.

Furthermore, physics, unlike mathematics, is generally seen to be rooted in
experiment and observations, and these fundamentally come down to conscious
experience.  This is particularly true in quantum mechanics, where in many
formulations observations are crucial.  If one refuses to consider conscious
observations, then there seems to be nothing wrong or lacking in bare QM,
though then it is just a beautiful mathematical theory for an unconscious world
that is totally divorced from observations.  However, precisely at the stage at
which one wants to explain what is consciously observed, one needs at least
some glimpse of a framework for connecting consciousness to observations.
Under the assumption that classical mechanics (CM) correctly described our
universe, many physicists usually imagined, I suspect, some simple idealized
form of psycho-physical parallelism in which the content of one's conscious
perception is very similar to certain aspects of the configuration of the
classical universe.  Then, say, if one looked at a clock whose hands were at
the 12 o'clock position, one would have a conscious perception that the clock
read 12 o'clock.

One of the problems of ordinary QM is that this simple form of psycho-physical
parallelism that seemed adequate in CM does not work in such a na\"{\i}vely
straightforward way.  Even though many physicists do not explicitly wish to
consider any theories of consciousness, I suspect that much of the trouble they
have with QM arises from the fact that they actually implicitly have something
like this simple form of psycho-physical parallelism sketched above, and then
they find difficulty fitting it with QM.  This is indeed one of the main
motivations for BM, since it has a definite trajectory, with a definite
position in configuration space (e.g., definite particle positions) at each
time, to which the simple form of psycho-physical parallelism can readily be
attached.  Since the sketch of a theory of consciousness is probably implicit
in many physicists' assumptions, one does not need to refer to it explicitly
when one extols the merits of a definite trajectory in BM, which one can thus
do without making other physicists uncomfortable by mentioning the subject of
consciousness that physicists do not really understand very well yet.

However, since Sensible Quantum Mechanics
[11-15]
gives a glimpse of a framework for a psycho-physical parallelism to the quantum
world that appears to be just as adequate (though of course just as sketchy in
detail) as the simple form implicitly assumed for a classical world, one can
see at least this one general possibility for overcoming the problem that many
physicists have in relating QM with conscious observations (a problem made
worse by a reluctance to consider consciousness explicitly and to acknowledge
one's na\"{\i}ve preconceptions about it).  SQM seemed extremely obvious when I
hit upon it, so initially it was surprising to me that it had not been
developed many years ago (though I did later find that Lockwood \cite{Lo} had a
few years ago expressed highly concordant ideas in less mathematical form).  It
occurred to me that the novelty of SQM was perhaps simply because of
physicists' reluctance to consider consciousness explicitly and because of the
strong psychological hold on their assumptions of the simple form of
psycho-physical parallelism possible for the classical world.

In view of the apparent success of SQM in sketching how QM may be combined with
consciousness with virtually none of the traditional interpretive problems (at
least in my own eyes; most physicists I have talked to still seem to think that
any theory of consciousness is either unnecessary or premature), it may be of
interest to develop also an SBM theory to combine BM in a similar way with
consciousness.  Then one can have two similar frameworks, containing
consciousness explicitly rather than merely implicitly as has unfortunately too
often been done, for comparing ordinary QM with BM.  Since in my nonstandard
view I consider SQM (which is the bare QM of an algebra of operators and a
quantum state, with no measurement hypothesis or collapse of the wavefunction,
augmented by a theory of consciousness connected to the bare QM by a
nonclassical form of psycho-physical parallelism) to be superior to all other
forms of ordinary QM, I shall not bother comparing my SBM extension of BM to
other forms of QM, but only to what in my biased opinion I consider to be the
best form of ordinary QM with consciousness, SQM.

Since BM is even at the unconscious level an augmentation of bare QM, it would
be simplest to describe first SQM and then show how it can be augmented to SBM.
 SQM is given by the following three basic postulates or axioms \cite{P95}:

 {\bf Quantum World Axiom}:  The unconscious ``quantum world'' $Q$ is
completely described by an appropriate algebra of operators and by a suitable
state $\sigma$ (a positive linear functional of the operators) giving the
expectation value $\langle O \rangle \equiv \sigma[O]$ of each operator $O$.

 {\bf Conscious World Axiom}:  The ``conscious world'' $M$, the set of all
perceptions $p$, has a fundamental measure $\mu(S)$ for each subset $S$ of $M$.

 {\bf Quantum-Consciousness Connection}:  The measure $\mu(S)$ for each set $S$
of conscious perceptions is given by the expectation value of a corresponding
``awareness operator'' $A(S)$, a positive-operator-valued (POV) measure
\cite{Dav}, in the state $\sigma$ of the quantum world:
 \begin{equation}
 \mu(S) = \langle A(S) \rangle \equiv \sigma[A(S)].
 \label{eq:1}
 \end{equation}

Here the Quantum World Axiom is the basic axiom of the mechanics of bare QM,
and the Conscious World Axiom is the basic postulate I shall make about the
conscious world, even when I go from SQM to SBM.  Perhaps I should remind the
reader \cite{P95} that a perception $p$ is in this context taken to mean the
entirety of a single conscious experience, all that one is consciously aware of
or consciously experiencing at one moment, the total ``raw feel'' that one has
at one time, or \cite{Lo} a ``phenomenal perspective'' or ``maximal
experience.''

Besides a modification of the Quantum World Axiom in BM, in SBM I shall also
modify the Quantum-Consciousness Connection, though in both SQM and SBM I shall
assume that the measure for each set of conscious perceptions is given by some
functional of the corresponding quantum or mechanical world.  In SQM the
measure is a functional of the operators and the quantum state, and I made the
simplest assumption that this functional is linear in the quantum state and is
the expectation value of a particular operator (an ``awareness operator''
$A(S)$) for each set $S$ of perceptions.  Of course, one could readily
contemplate generalizations in which the measure is a {\it nonlinear}
functional of expectation values \cite{P95}, but I shall not do that here.

Bohmian mechanics is most simply given in the case in which a quantum state is
given by a time-dependent wavefunction over some configuration space (say, for
simplicity, of particle positions).  One augments this quantum state by a
history or time-parametrized trajectory in this configuration space with the
velocity vector at each point being given by a suitable functional of the
wavefunction (a normalized gradient of its phase for spinless particles)
[1-10].
This gives a first-order differential equation for the trajectory, so it is
uniquely determined by this equation (which is itself uniquely determined by
the wavefunction except at its zeros, but those isolated points do not cause
any trouble) and by the position of the trajectory at any one (e.g., initial)
time.

Because BM thus has a trajectory as well as a quantum state, there are in
principle more things on which the measure for conscious perceptions can
depend.  Again assuming a linear (if nontrivial) dependence on the quantum
state, I shall propose that Sensible Bohmian Mechanics (SBM) is given by the
following three axioms:

 {\bf Bohmian World Axiom}:  The unconscious ``Bohmian world'' $B$ is described
by a normalized time-dependent Hilbert-space wavefunction $\psi(x,t)$ (obeying
an appropriate Schr\"{o}dinger equation that determines the wavefunction at all
times $t$ once it is given at one time) on a configuration space with
coordinates $x$ (indices suppressed) and by a Bohmian trajectory $T$ in this
configuration space whose velocity is given by a suitable functional of the
wavefunction.

 {\bf Conscious World Axiom}:  The ``conscious world'' $M$, the set of all
perceptions $p$, has a fundamental measure $\mu(S)$ for each subset $S$ of $M$.

 {\bf Bohmian-Consciousness Connection}:  The measure $\mu(S)$ for each set $S$
of conscious perceptions is given by the expectation value of a corresponding
``awareness operator'' $A(S,T)$, a positive-operator-valued (POV) measure
\cite{Dav} that depends on the Bohmian trajectory $T$, in the wavefunction of
the Bohmian world:
 \begin{equation}
 \mu(S) = \int{dx \psi^*(x,t_0) A(S,T) \psi(x,t_0)}.
 \label{eq:2}
 \end{equation}
(Here the integral can be taken at any time $t_0$, since the Schr\"{o}dinger
equation determines the wavefunction at all times from its form at time $t_0$,
though of course the form of $A(S,T)$ would depend on the particular $t_0$
chosen.  Also, for simplicity, I am suppressing possible spin indices on the
wavefunction as well as the coordinate indices on the $x$ that denotes a point
in the multi-dimensional configuration space.)

As in SQM \cite{P95}, so in SBM it is convenient to hypothesize that the set
$M$ of all possible conscious perceptions $p$ is a suitable topological space
with a prior measure
 \begin{equation}
 \mu_0(S) = \int_{S}{d\mu_0(p)}.
 \label{eq:3}
 \end{equation}
Then, just as for SQM, the linearity of positive-valued-operator measures over
sets allows one to write the awareness operators for SBM as
 \begin{equation}
 A(S,T) = \int_S E(p,T)d\mu_0(p),
 \label{eq:4}
 \end{equation}
a generalized sum or integral of SBM ``experience operators'' or ``perception
operators'' $E(p,T)$ for the individual perceptions $p$.  Similarly, one can
write the measure on a set of perceptions $S$ as
 \begin{equation}
 \mu(S) = \int_S d\mu(p) = \int_S m(p) d\mu_0(p),
 \label{eq:5}
 \end{equation}
in terms of a measure density $m(p)$ that is the quantum expectation value of
the experience operator $E(p,T)$ for the same perception $p$:
 \begin{equation}
 m(p) = \int{dx \psi^*(x,t_0) E(p,T) \psi(x,t_0)}.
 \label{eq:6}
 \end{equation}

It is simplest to consider the two extreme cases in which, for each $S$ or $p$,
$\mu(S)$ or $m(p)$ depends only on the wavefunction $\psi(x)$ or on $T$.  In
the first of these cases, one essentially has a form of SQM with the quantum
world consisting of a state that has the particular form of a time-dependent
Hilbert-space wavefunction on some configuration and of operators on such
wavefunctions, and with the Bohmian world consisting of this quantum world
augmented by the Bohmian trajectory $T$.  However, in this case the Bohmian
trajectory has absolutely no effect on the conscious world with its perceptions
and measures.  Since our perceptions are the only direct contact we have with
the world, if they are completely unaffected by the Bohmian trajectory, there
would then seem to be no motivation to add them to one's theory of the quantum
world.  Although it indeed fits my own prejudice not to augment bare QM for the
unconscious aspects of the world and thus not to bother with any Bohmian
trajectory in theories of the quantum world that I prefer, for the purpose of
the present discussion on BM and its extension to SBM it would seem better to
consider a Bohmian-Consciousness Connection in which the Bohmian trajectory
really does have an effect on the measure for sets of perceptions.  As
discussed above, a motivation for the trajectory in BM is that its point in
configuration space at each time seems closer to the content of conscious
perceptions than does the quantum state or wavefunction itself.

The second extreme possibility is that for each $S$ or $p$, $\mu(S)$ or $m(p)$
depends only on the trajectory $T$.  This can be accomplished in
Eqs.~(\ref{eq:2}) or (\ref{eq:6}) by having $A(S,T)$ or $E(p,T)$ not have any
nontrivial dependence on $x$ but simply be an $S$- or $p$-dependent numerical
function purely of $T$, say $A(S,T)=a(S,T)$ (i.e., a function purely of $S$ and
of $T$ times the identity operator in the wavefunction Hilbert space, so for a
normalized wavefunction, one gets $\mu(S)=a(S,T)$), or, say, $E(p,T)=e(p,T)$
(so then $m(p)=e(p,T)$).

In this case one might say that he could simply dispense with the wavefunction
$\psi(x,t)$, since the measure for one's perceptions would then depend only on
the Bohmian trajectory $T$.  However, the time dependence of this trajectory
depends in a simple way upon the wavefunction, so the latter could still be a
useful element in giving the simplest description of the trajectory upon which
the measure for perceptions depends.  (Similarly, one could in principle
dispense with the quantum or Bohmian world altogether and consider merely the
existence of the conscious world, or actually even only one's own present
conscious perception within it, as that is all he has direct experience of, but
the description of even just one's own present conscious perception by itself
may be simpler if one postulates a whole conscious world with many measured
sets of perceptions, and a description of this conscious world by itself may
also be simpler if one postulates the existence of a quantum or Bohmian world
on which it depends.  I personally also believe that a description of the full
SQM or SBM world would be simpler if one postulates the existence of an
omniscient, omnipotent God as the Creator of this world, but of course this
further extrapolation from direct conscious experience takes one beyond what is
traditionally called physics to metaphysics.)

If for each $S$ and $p$, $A(S,T)$ and $E(p,T)$ are the purely numerical
functions (or functionals) $a(S,T)$ and $e(p,T)$ of the trajectory $T$, with no
dependence on the wavefunction, there is still the question of what kind of
functions of $T$ they are.  In principle they could be nonlocal functionals of
the entire trajectory, so that one's perception depended on the entire Bohmian
history.  The fact that we do not normally perceive much detail about the
future might be taken to suggest that there is a time $t(p)$ associated with
each perception, and that $e(p,T)$ depends only on the part of the trajectory
at times at and earlier than $t(p)$.  The dependence on the part of $T$ at
times earlier than $t(p)$ might then be postulated to give the memory
components of perceptions.

However, the fact that drugs and physical conditions such as Alzheimer's
disease seem to have a strong effect on memory suggests to me that it is more
plausible to assume that the memory components of present perceptions are
directly caused by present physical conditions (perhaps having something to do
with neural connections in the brain) rather than being caused directly by past
conditions such as the past part of the Bohmian trajectory.  Thus even if the
measure density $m(p)$ for a perception $p$ were determined entirely by the
Bohmian trajectory $T$ rather than by it and by the quantum state (or entirely
by the quantum state, as in SQM, which I personally prefer), I would think it
to be the simplest plausible assumption that $m(p)$ is determined by the
trajectory at the single time $t(p)$, which can be considered to be the time of
the perception (something not well defined in SQM except in rather {\it ad hoc}
ways \cite{P95}).  In other words, I am claiming that it seems simplest to
suppose that each conscious perception is directly only of the present, though
that present can include the records that produce the perception components of
memory.

Of course, it may be an extreme idealization to suppose that the dependence is
on precisely that single time rather than being spread out over, say, one
Planck time.  But I see no evidence that each of our perceptions need be
directly affected by anything in the unconscious quantum or Bohmian world
spread out over a time long in, say, seconds.  Although a second is admittedly
very long compared with the Planck time, it is certainly much shorter than
cosmological times, and there is no evidence that I see that puts a lower limit
on the time over which the unconscious quantum or Bohmian world affects a
single conscious perception (unless one adopts a theory in which instants and
time periods less than some lower limit simply do not exist).

Even if the measure density $m(p)$ for a perception $p$ is determined entirely
by the Bohmian trajectory at the instant $t(p)$, there is still the question of
what aspect of the trajectory at that instant determines $m(p)$.  In principle,
it could be determined by the position, velocity, acceleration, and/or higher
time derivatives of the position in the configuration space $x$ as a function
of time $t$, evaluated at the time $t(p)$.

If one wants to get the closest agreement between an SQM theory and a version
of SBM with $m(p)$ depending purely on the Bohmian trajectory, it seems that
one should select the following features in one's choice of the SQM and the
SBM:

(1)  The SQM and SQM should have the same bare quantum world, a normalized
time-dependent Hilbert-space wavefunction on some configuration space.

(2)  The position $x_0$ of the Bohmian trajectory $T$ at time $t_0$ should be
chosen randomly with the probability distribution given by the quantum
``probability'' density $\psi^*(x,t_0)\psi(x,t_0)$.  (The Bohmian equation for
the velocity of the trajectory as the gradient of the phase of the wavefunction
ensures that if one selects a continuum ensemble of trajectories such that at
time $t_0$ their measure density in the configuration space is the quantum
``probability'' density $\psi^*(x,t_0)\psi(x,t_0)$ at that time, then the
measure density of this ensemble of trajectories is carried forward in time
such that at time $t$ it is precisely the quantum ``probability'' density
$\psi^*(x,t)\psi(x,t)$ at that new time $t$.  In this sense the ``random''
probability distribution $\psi^*(x,t)\psi(x,t)$ preserves its functional
dependence on the wavefunction $\psi(x,t)$ at all times.)

(3) The SQM should have each of its experience operators $E(p)$ being a
projection operator $P(s(p),t(p))$ onto a subset $s(p)$ of the configuration
space and at a time $t(p)$ that both depend on the perception $p$.

(4)  The SBM should have each of its experience operators being of the form
$E(p,T)=e(p,T)$ and having a nonzero constant value (say 1) if the trajectory
$T$ at the time $t(p)$ (the same function of the perception $p$ as in the
corresponding SQM theory) is within the subset $s(p)$ of the configuration
space (the same subset for each $p$ as in the corresponding SQM theory), and
having the value zero if the trajectory at time $t(p)$ is not within $s(p)$.

Even in this idealized correspondence between SQM and SBM theories, one will
not get for each the same measure density $m(p)$ for individual perceptions or
the same measure $\mu(S)$ for sets of perceptions.  However, if one enlarged
the BM and SBM theories to {\it Continuum Bohmian Mechanics} (CBM) and {\it
Sensible Continuum Bohmian Mechanics} (SCBM) theories respectively that are
hereby defined to include a whole continuum ensemble of trajectories, with
their measure density given by $\psi^*(x,t)\psi(x,t)$ at any time $t$ (whose
preservation at all times is guaranteed by the Bohmian equation of evolution
for each trajectory in the ensemble), then one would get agreement if one
integrated, over the measured continuum ensemble of all trajectories, the
$m(p)$, and hence also the $\mu(S)$, that one gets from each trajectory in the
particular SBM theory with its own individual trajectory.  In contrast, for a
particular SBM theory with its own particular trajectory $T$, one would get a
zero measure for any set $S$ of perceptions $p$ which all give
configuration-space subsets $s(p)$ that the particular trajectory $T$ is not in
at the corresponding times $t(p)$.  (On the other hand, any set of perceptions
with nonzero measure in any particular SBM theory would necessarily have a
nonzero measure in the corresponding SQM theory when features (1)-(4) hold for
the pair of corresponding theories.)

Nevertheless, the fact that each perception $p$ that actually occurs (i.e.,
that has a nonzero measure density) has no direct awareness of any other
perception means that one cannot absolutely test whether or not one is in an
SBM universe in which another perception $p'$, with the particular trajectory
$T$ not in the configuration-space subspace $s(p')$ at the time $t(p')$ and
hence with $m(p')=0$, does not actually occur, or whether one is in the
corresponding SCBM or SQM universe in which $m(p') > 0$ so that this other
perception $p'$ does actually occur.  It seems that perhaps the best one can do
is to compare what the corresponding SBM and SQM theories give for the {\it
typicality} of the perception $p$ (the fraction of the measure of other
perceptions $p'$ that have a measure density not greater than that of $p$, or
some modification of this such as the {\it dual typicality} \cite{P95}).
Although if the features (1)-(4) proposed above hold, the corresponding SCBM
and SQM theories would give the same typicality for each perception, even the
typicality for an existing perception in the SBM theory and that for the same
perception in the corresponding SQM theory would differ.  However, one might
hope that for a reasonably large fraction of individual Bohmian trajectories,
chosen randomly out of the continuum ensemble with the $\psi^*(x,t)\psi(x,t)$
measure density, most of the measure of the actually existing perceptions in
the resulting particular SBM theory would occur for a set of actually existing
perceptions whose typicalities would not be too low in either that particular
SBM theory or in the corresponding SQM theory.  If this were true (and its
truth might well depend on the form of the underlying bare quantum theory,
i.e., on the wavefunction, so that this could well be a question worth further
investigation), then the typicality of most existing perceptions would not give
a strong test between a particular SBM theory that predicted its existence and
the corresponding SQM theory.

The fact that one's perception $p$ cannot absolutely rule out any SBM or SQM
theory that gives it a nonzero measure density $m(p)$, and the fact that one
can apparently only use something like the typicality of the perception
(interpreted as the likelihood or conditional probability for the perception
given the theory) to weight the prior probability assigned to the theory in a
Bayesian analysis to get the posterior probability for the theory \cite{P95},
means that even if one has a pair of SQM and SBM theories that do not share the
features (1)-(4) above, one's actual perception will not necessarily rule out
either of these theories.  But, crudely speaking, the greater the degree to
which the features (1)-(4) are not held by the pair of theories, the greater
the possibility apparently is that the typicalities assigned by the two
theories to one's perception will differ sufficiently greatly that one can use
them to deduce a very low posterior probability to one or the other of the pair
of theories.

Of course, I should emphasize that the ``one'' who is postulated to be doing
the comparison of typicalities must be one who can indeed calculate them from
the SQM and SBM theories in question.  In practice this might be possible only
for a being of such intelligence that he, she, or it (or whatever is the
correct pronoun for such a being that may not have any sex and yet is more
intelligent than we usually ascribe to ``things'') can exist only outside our
universe.  Those of us within the universe might be expected to be able to get
only a very crude estimate of such typicalities, but if one can even see
roughly that they differ by many orders of magnitude between two theories, that
would be sufficient to get a reasonably good idea of which theory to reject
(except in the case in which the prior probabilities differed by roughly the
same number of orders of magnitude in the opposite direction).

There is also the apparently completely subjective question of what prior
probabilities to assign different SQM and SBM theories before weighting them
with the typicalities of one's perception in a Bayesian analysis to get the
posterior probabilities for the theories.  (I say ``apparently,'' since
conceivably there is an actual existing measured set of different universes,
each described by a different SQM or SBM theory, so that the actual measure of
these universes gives an ontologically objective prior frequency-type
probability to the different theories describing the different universes, but
since we certainly do not have access to all of these conceivable universes or
know their measure even if it does exist objectively, for us epistemologically
the prior measure must surely be subjective.)  Here I should lay my cards on
the table and explain the prejudices that I have against assigning BM and SBM
theories high prior probabilities.

First, I should say that I would prefer to assign higher prior probabilities to
simpler theories.  Perhaps most physicists would agree, but they might differ
on how to weight or even how to rank the simplicity of different theories.  (I
suspect that this may lie at the core of more metaphysical disagreements as
well, such as whether theism or atheism is true.)

I myself think that adding the Bohmian trajectory to bare quantum mechanics
reduces the simplicity and thus the prior probability I would assign to the
theory.  Perhaps for nonrelativistic QM, the extension to CBM with its
continuum of trajectories that have the $\psi^*(x,t)\psi(x,t)$ measure density
is not too great a loss of simplicity, since this ensemble of trajectories does
not take too much additional information to specify and has certain nice
properties (such as the measure density's remaining $\psi^*(x,t)\psi(x,t)$ for
all time).  However, to pick out a single trajectory from this ensemble for a
particular BM (as opposed to CBM) theory would seem to require much more {\it
ad hoc} information.  I can think of certain choices that do not require too
much additional information, such as choosing the trajectory whose time
integral of $\psi^*(x,t)\psi(x,t)$ evaluated along the trajectory itself is the
maximum out of all possible Bohmian trajectories, but even these relatively
simple choices seem rather {\it ad hoc} and ugly.

Incidentally, I might point out that one could have theories of {\it
Generalized Continuum Bohmian Mechanics} (GCBM), and its augmentation of {\it
Sensible Generalized Continuum Bohmian Mechanics} (SGCBM) to include the
Conscious World Axiom and the (suitably generalized, as discussed above for
SCBM) Bohmian-Consciousness Connection, in which there is a continuum of
Bohmian trajectories with a measure density that is a general normalized
nonnegative function over the configuration space at any one time and which is
transported to other times by the Bohmian equation for the velocity of the
trajectories.  One could then say that CBM is the special case (probably the
simplest) in which this measure density is $\psi^*(x,t)\psi(x,t)$, and ordinary
BM with its single trajectory is the special case in which the measure density
at any one time is a delta-function distribution (not so simple, as discussed
above, since one must specify the location of this delta-function at one time.)
 However, even within the wide set of possibilities of GCBM or SGCBM theories,
I do not see any so simple as simply leaving out the trajectories altogether,
as in bare QM and SQM theories.

Now a supporter of BM might object that although BM is almost certainly more
complicated than bare QM in requiring the extra element of the Bohmian
trajectory, in another way it is simpler in not requiring any operators on the
Hilbert space of wavefunctions (except presumably for the Hamiltonian that
generates the evolution by the Schr\"{o}dinger equation).  However, once the
Hilbert space of wavefunctions is defined, it takes very little additional
information to define the set of operators on this space that maps it into
itself.  Perhaps this additional information is comparable to that for the
continuum ensemble of trajectories in CBM, so I would admit that CBM is of
comparable simplicity to bare QM for the same Hilbert space (though specifying
a precise single trajectory in ordinary BM seems to me in general more
complicated).  But even in a comparison with CBM, the framework of bare QM
allows a much greater range of possibilities for the quantum state than the
restriction to a wavefunction over configuration space (or some slight
generalizations to include spin, etc.) that seems to be necessary for present
formulations of BM and CBM, and it might turn out that a quantum state that
cannot be written as a wavefunction will be at the heart of the simplest
complete description of the universe.

A supporter of BM might also object that even if BM is indeed more complicated
than bare QM in having its Bohmian trajectory, when one attaches a theory of
consciousness, the result is simpler if one can attach it to a theory with a
trajectory.  For example, in the Sensible theories outlined above, a supporter
of SBM over SQM might say that BM allows simpler awareness operators $A(S,T)$
(e.g., of the form $a(S,T)$, with no nontrivial dependence on the wavefunction)
in the Bohmian-Consciousness Connection of SBM than the awareness operators
$A(S)$ allowed in the Quantum-Consciousness Connection of SQM, and that the
increased simplicity of  $A(S,T)$ over $A(S)$ overbalances the extra
complication of the trajectory in SBM.  I will admit that it is hard to answer
this objection when we are as ignorant as we are about the connection of
consciousness to the rest of physics (e.g., about what the awareness operators
are in Sensible theories), but I should say that at present I for one am rather
sceptical that a viable SBM theory will turn out to be as simple as a viable
SQM theory.

Another problem I have with BM theories is that when I imagine applying them to
relativistic fields instead of to nonrelativistic particles, I find that even
if the quantum state is Lorentz invariant (e.g., the vacuum state), almost all
the trajectories are not, and neither is the continuum ensemble of CBM.  For
example, consider for simplicity a free scalar field.  Each plane-wave mode of
the field can be considered to be an harmonic oscillator, and the vacuum state
of the field can be considered to be a product of the ground states of all of
these modes.  For such a static state of zero energy, the wavefunction has a
constant phase with zero gradient in the configuration space (the space of
amplitudes of the modes), so the Bohmian trajectory is static.  When one
superposes the static amplitudes for the modes, one gets an arbitrary
space-dependent static configuration of the scalar field as its Bohmian
trajectory.  Except for the special homogeneous cases in which this static
configuration is precisely the configuration of constant field everywhere
(which in the case of zero field is the one proposed above whose time integral
of $\psi^*\psi$ evaluated along the trajectory itself is the maximum out of all
possible Bohmian trajectories), these Bohmian trajectories are not Lorentz
invariant, for the spatial dependence in the frame in which their evolution was
calculated by the Bohmian equation of motion becomes a temporal dependence in
any other frame.

Furthermore, in any other frame the trajectory which does not even satisfy the
Bohmian equation of motion, so the situation is worse than the version of
spontaneous symmetry breaking in which the quantum state is not invariant under
the symmetry group of the equations which it solves.  In that case the action
of an element of the symmetry group on the nonsymmetric solution yields another
nonsymmetric solution, but for the inhomogeneous Bohmian trajectories, the
action of a Lorentz transformation on them yields time-dependent trajectories
that do not even satisfy the Bohmian equations for trajectories.  (I am
grateful to Shelly Goldstein for pointing out the importance of this
distinction.)

Of course, one could say that perhaps Lorentz invariance is just a useful
approximation to certain aspects of the world and need not apply to the Bohmian
trajectories.  After all, in our perceptions there are objects that seem to
have fairly definite velocities (with the bulk of nearby ones, such as the
earth, being fairly near zero in our local frame) that appear to break Lorentz
invariance, so presumably there is nothing blatantly inconsistent with
observations to have Bohmian trajectories breaking Lorentz invariance even when
the quantum state is Lorentz invariant.  However, this lack of Lorentz
invariance for all but the homogeneous Bohmian trajectories for a quantum field
in the Lorentz-invariant vacuum state, and the need to define a preferred
velocity in order to define the Bohmian equations for the trajectories, seems
at least aesthetically rather ugly.

I suspect that there would be an even greater degree of ugliness if one
attempted to devise a Bohmian version of quantum gravity, say for the quantum
cosmology of closed universes.  If one has a solution of the Wheeler-DeWitt
equation for canonically-quantized general relativity, this gives a
wavefunctional of three-geometries that is invariant not only under coordinate
transformations of the spatial hypersurface, but also under the equivalent of
time translations that are arbitrary at each point of space.  Unlike the case
of quantum field theory in Minkowski spacetime, in which only one quantum state
(the vacuum) has the full Lorentz invariance of the Minkowski spacetime itself,
in canonical quantum gravity {\it every} physical state has the diffeomorphism
invariance (including local Lorentz invariance) of general coordinate
transformations.  It is not clear to me how to construct a Bohmian trajectory
for such a theory of quantum cosmology without breaking this diffeomorphism
invariance in a complicated and ugly {\it ad hoc} way.

I would also suspect that if one tries to overcome the severe technical
problems of constructing a finite theory of gravity by going from a field
theory of general relativity to a superstring theory, it would be even more
cumbersome and ugly to try to construct a Bohmian trajectory.  Of course, the
bare QM of superstrings at the nonperturbative level is hardly understood at
the moment, and virtually nothing is known of how to produce consciousness out
of it (say by some form of awareness operators), so one cannot pretend to be
certain that this task will be easier without constructing a BM version of
superstrings, but I personally suspect that the latter would be simply an
onerous burden that is unnecessary.  It would seem much more plausible that one
should go directly from a bare quantum theory of superstrings (or whatever the
ultimate quantum theory is, assuming that it {\it is} a quantum theory) to a
theory of consciousness (say given in terms of the unconscious quantum world by
expectation values of awareness operators in SQM), without bothering with
trying to find a BM theory with a definite trajectory.

In conclusion, I have outlined how a theory of consciousness may be attached to
the Bohmian version of quantum mechanics (BM) to give what I call Sensible
Bohmian Mechanics (SBM), in a way highly analogous to the way in which I have
proposed attached a theory of consciousness to ordinary quantum mechanics (QM)
to give Sensible Quantum Mechanics (SQM)
[11-15].  Because BM has a Bohmian trajectory that bare QM does not have, SBM
for a fixed wavefunction has more possibilities for awareness operators for
giving the measure of sets of conscious perceptions.  (Conversely, BM seems to
require a wavefunction on a configuration space, or something like it, which QM
does not, so in the latter way QM is more flexible.)  However, if a SBM is to
give mostly typical perceptions that would also be typical in a corresponding
SQM theory, it seems that there should be at least four features of the
correspondence (listed above), though it is not completely clear that these are
entirely sufficient.  Although SBM might conceivably allow a simpler set of
awareness operators than SQM, this is by no means obvious, and in other
respects SBM seems to me to be more complicated and ugly than SQM.

I especially appreciate many helpful discussions with Shelly Goldstein, though
he does not share my subjective negative evaluation of Bohmian mechanics.
Financial support has been provided by the Natural Sciences and Engineering
Research Council of Canada.


\begin{thebibliography}{99}

\bibitem{deB} L. de Broglie, ``La nouvelle dynamique des quanta,'' in {\em
Electrons et Photons: Rapports et
Discussions du Cinqui\`eme Conseil de Physique tenu \`a Bruxelles du 24 au 29
Octobre 1927 sous les Auspices de l'Institut International de Physique Solvay}
(Gauthier-Villars, Paris, 1928), pp. 105-132; {\em Physicien et Penseur}
(Paris, 1953); {\em Tentative d'Interpr\'{e}tation Causale et
Non-lin\'{e}aire de la M\'{e}canique Ondulatoire} (Gauthier-Villars, Paris,
1956); Found.\ Phys.\ {\bf 1}, 5 (1970).

\bibitem{Bohm} D. Bohm, Phys.\ Rev.\ {\bf 85}, 166 and 180 (1952), reprinted in
{\em Quantum Theory and Measurement}, edited by J. A. Wheeler and W. H. Zurek
(Princeton University Press, Princeton, 1983), p. 369 and 383; Phys.\ Rev.\
{\bf 89}, 458 (1953).

\bibitem{BH} D. Bohm and B. J. Hiley, {\em The Undivided Universe:  An
Ontological Interpretation of Quantum Theory} (Routledge and Kegan Paul,
London, 1993).

\bibitem{Be} J. S. Bell, {\em Speakable and Unspeakable in Quantum Mechanics}
(Cambridge University Press, Cambridge, 1987).

\bibitem{DGZ} D. D\"{u}rr, S. Goldstein, and N. Zanghi, J.\ Stat.\ Phys.\ {\bf
67}, 843 (1992); Found.\ Phys.\ {\bf 23}, 721 (1993); Phys.\ Lett.\ {\bf A172},
6 (1992).

\bibitem{A92} D. Z. Albert, {\em Quantum Mechanics and Experience} (Harvard
University Press, Cambridge, Massachusetts, 1992).

\bibitem{Ho} P. R. Holland, {\em The Quantum Theory of Motion} (Cambridge
University Press, Cambridge, 1993).

\bibitem{C} J. T.  Cushing, {\em Quantum Mechanics:  Historical Contingency and
the Copenhagen Hegemony} (University of Chicago Press, Chicago, 1994).

\bibitem{A94} D. Z. Albert, Sci.\ Am.\ {\bf 270}, 58 (May 1994).

\bibitem{BDDGZ} K. Berndl, M. Daumer, D. D\"{u}rr, S. Goldstein, and N. Zanghi,
``A Survey of Bohmian Mechanics'' (University of Munich report, April 1995),
quant-ph/9504010.

\bibitem{P94a} D. N. Page, ``Probabilities Don't Matter,'' to be published in
{\em Proceedings of the 7th Marcel Grossmann Meeting on General Relativity},
edited by M. Keiser and R. T. Jantzen (World Scientific, Singapore 1995)
(University of Alberta report Alberta-Thy-28-94, Nov. 25, 1994), gr-qc/9411004.

\bibitem{P94b} D. N. Page, ``Information Loss in Black Holes and/or Conscious
Beings?'' to be published
in {\em Heat Kernel Techniques and Quantum Gravity}, edited by S. A. Fulling
(Discourses in Mathematics and Its Applications, No. 4, Texas A\&M University
Department of Mathematics, College Station, Texas, 1995) (University of Alberta
report Alberta-Thy-36-94, Nov. 25, 1994), hep-th/9411193.

\bibitem{P95} D. N. Page, ``Sensible Quantum Mechanics:  Are Only Perceptions
Probabilistic?'' (University of Alberta
report Alberta-Thy-05-95, June 7, 1995), quant-ph/9506010.

\bibitem{P95c} D. N. Page, ``Sensible Quantum Mechanics:  Are Probabilities
Only in the Mind?'' to be published in {\em Proceedings of the Sixth Seminar on
Quantum Gravity, June 12-19, 1987, Moscow, Russia}, edited by V. A. Berezin and
V. A. Rubakov (World Scientific, Singapore, 1996) (University of Alberta report
Alberta-Thy-13-95, July 4, 1995), gr-qc/9507024.

\bibitem{P95d} D. N. Page, ``Aspects of Quantum Cosmology,'' to be published in
{\em Proceedings of the International School of Astrophysics ``D. Chalonge,''
4th Course:  String Gravity and Physics at the Planck Energy Scale, Erice,
Sicily, 8-19 September 1995}, edited by N. Sanchez (Kluwer, Dordrecht, 1996)
(University of Alberta report Alberta-Thy-14-95, July 10, 1995), gr-qc/9507025.

\bibitem{Lo} M. Lockwood, {\em Mind, Brain and the Quantum:  The Compound `I'}
(Basil Blackwell, Oxford, 1989).

\bibitem{Dav} E. B. Davies, {\em Quantum Theory of Open Systems} (Academic
Press, London, 1976).

\end{thebibliography}
\end{document}